\documentclass[10pt, conference]{IEEEtran}

\usepackage{graphicx}
\usepackage{subfigure}
\usepackage{float}
\usepackage{amsmath}
\usepackage{amsfonts,amssymb}
\usepackage{dsfont}
\usepackage{algpseudocode}
\usepackage{makecell}
\usepackage{cite}
\usepackage{multirow}
\usepackage{multicol}
\usepackage{booktabs}
\usepackage[ruled,vlined, linesnumbered,lined,boxed,commentsnumbered,ruled,longend,noend]{algorithm2e}
\usepackage{xcolor,soul}
\usepackage{comment}
\usepackage{bbm}
\usepackage{tabularx}
\usepackage[flushleft]{threeparttable}
\usepackage{nicefrac}
\usepackage[linesnumbered,ruled,vlined]{algorithm2e}
\usepackage{fancyhdr,lipsum}

\SetKwInput{KwInput}{Input}                
\SetKwInput{KwOutput}{Output}              

\fancypagestyle{mahmood}{%
   \fancyhf{} 
   
   \fancyhead[C]{\tiny This manuscript has been accepted for presentation at \textbf{IEEE GLOBECOM 2025}. You can use this material personally. Reprinting or republishing this material for the purpose of advertising or promotion, creating new collective works, reselling or redistributing to servers or lists, or using any copyrighted component in other works \textbf{must adhere to IEEE policy}. The DOI will be supplied as soon as it becomes available.}
}%

\makeatletter
\let\ps@IEEEtitlepagestyle\ps@mahmood
\makeatother

\begin{document}
\title{Adaptive Multiple Access and Service Placement for Generative Diffusion Models}

\author{
    \IEEEauthorblockN{
        Hamidreza Mazandarani\textsuperscript{1}, Mohammad Farhoudi\textsuperscript{2}, Masoud Shokrnezhad\textsuperscript{3},
        and Tarik Taleb\textsuperscript{1} \\
    }
    \IEEEauthorblockA{
       \textsuperscript{1} \textit{Ruhr University Bochum, Bochum, Germany; \{ hamidreza.mazandarani, tarik.taleb\}@rub.de} \\
        \textsuperscript{2} \textit{Oulu University, Oulu, Finland; mohammad.farhoudi@oulu.fi} \\
        \textsuperscript{3} \textit{ICTFICIAL Oy, Espoo, Finland}; masoud.shokrnezhad@ictficial.com \\
    }
}

\maketitle

\begin{abstract}
Generative Diffusion Models (GDMs) have emerged as key components of Generative Artificial Intelligence (GenAI), offering unparalleled expressiveness and controllability for complex data generation tasks. However, their deployment in real-time and mobile environments remains challenging due to the iterative and resource-intensive nature of the inference process. Addressing these challenges, this paper introduces a unified optimization framework that jointly tackles service placement and multiple access control for GDMs in mobile edge networks. We propose LEARN-GDM, a Deep Reinforcement Learning-based algorithm that dynamically partitions denoising blocks across heterogeneous edge nodes, while accounting for latent transmission costs and enabling adaptive reduction of inference steps. Our approach integrates a greedy multiple access scheme with a Double and Dueling Deep Q-Learning (D3QL)-based service placement, allowing for scalable, adaptable, and resource-efficient operation under stringent quality of service requirements. Simulations demonstrate the superior performance of the proposed framework in terms of scalability and latency resilience compared to conventional monolithic and fixed chain-length placement strategies. This work advances the state of the art in edge-enabled GenAI by offering an adaptable solution for GDM services orchestration, paving the way for future extensions toward semantic networking and co-inference across distributed environments.
\end{abstract}


\section{Introduction}\label{S_INT}

\begin{figure}[t!]\centering
\vspace{0.1cm}
\includegraphics[width=3.4in]{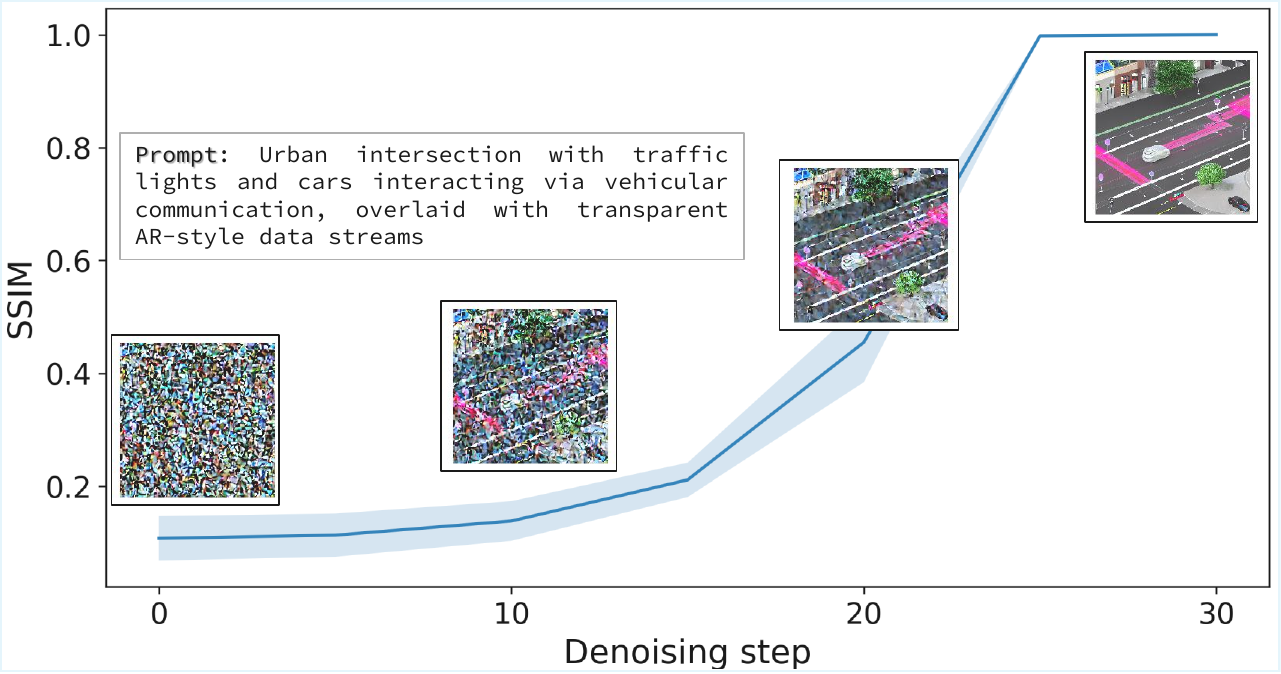}
\vspace{-0.35cm}
\caption{An example of GDM inference utilizing Stable Diffusion. The image quality, quantified by the Structural Similarity Index Measure (SSIM), gradually enhances during the denoising process. The shaded area illustrates the standard deviation computed across 10 distinct prompts.}
\vspace{-0.55cm}
\label{figure:SSIM_sample}
\end{figure}

Generative Diffusion Models (GDMs) have recently emerged as a compelling framework within Generative Artificial Intelligence (GenAI), enjoying the distinctive capability of generating high-quality outputs by progressively denoising stochastic input data \cite{10419041} (Fig. \ref{figure:SSIM_sample}). Their inherent expressiveness and controllability make them particularly suitable for complex generative tasks, leading to their integration into communication systems for tasks such as channel-distortion-aware image reconstruction \cite{letafati2024conditional}, image manipulation \cite{Salar_2025_CVPR}, and intelligent resource orchestration \cite{du2024enhancing, luong2025diffusion}. Furthermore, GDMs are being recently explored for diffusion-based reasoning \cite{ye2024diffusion}.

In vehicular networks, one of their applications involves the improvement of road intelligence to facilitate immersive in-vehicle experiences, which include the generation of real-time three-dimensional content \cite{xie2024gai}.
Despite their versatility, GDM services present significant challenges for real-time and mobile deployments due to their computationally intensive inference process arising from their iterative nature \cite{10419041}. Addressing this bottleneck has emerged as a central theme in recent literature, with approaches that include sampling step reduction and model compression, in conjunction with auto-regressive techniques \cite{tang2024hart}, as well as more systemic solutions that leverage the substantial computing power of edge nodes \cite{feng2024exploring, yang2024efficient, zeng2025generative, liu2025qos, li2024flexgen}.

In the latter category, the inherent gradual nature of the denoising process is acknowledged, and edge computing is seen as a promising enabler, offering low-latency access to computational resources for mobile users. In this regard, Feng \textit{et al.} \cite{feng2024exploring} proposed an edge-user collaborative GDM inference framework in which a proportion of the denoising steps can be offloaded to the edge servers. Also, Quality of Experience (QoE) for users is considered, which is defined as image quality minus weighted latency and energy consumption. Xie \textit{et al.} \cite{xie2024gai} introduced a framework for collaborative fine-tuning and distributed inference in vehicular networks, with a splitting strategy of inferences to optimize latency and content-generation capability. Similarly, in the proposal of Yang \textit{et al.} \cite{yang2024efficient}, the inference process for each user was split into two phases with an optimizable split point: a shared model for low-level generation at the edge, followed by personalized user models. Zeng \textit{et al.} \cite{zeng2025generative} adopted a different approach by partitioning multi-modal content and offloading partial diffusion tasks to multiple servers. Lie \textit{et al.} \cite{liu2025qos} proposed a reinforcement learning algorithm that leverages diffusion models and context-aware attention to improve multi-type task orchestration at the network edge. Finally, FlexGen \cite{li2024flexgen} sought to improve quality and cost adjustability by modulation of model width (i.e., the layer size).

Nevertheless, these efforts often fall short of addressing the dynamic and mobile nature of real-world users navigating through networks \cite{mazandarani2023self, shokrnezhad2025autonomous}. Particularly, the potential for distributing the denoising steps of GDM services across various heterogeneous edge nodes, while considering the latent transmission costs and adaptable chain lengths, remains underexplored. In addition, existing works do not account for multiple access policies that coordinate how users share transmission channels to avoid collisions and ensure fairness under resource constraints \cite{shokrnezhad2024fairness, mazandarani2025novel}. Building on our prior work on joint design of communications and computation \cite{shokrnezhad2024towards, farhoudi2024discovery, mazandarani2024semantic, globecom2023}, we propose an adaptable framework that jointly optimizes service placement and multiple access control for GDMs in mobile edge networks. We consider a group of Base Stations (BSs) equipped with computing resources, where mobile users intermittently request GDM-generated outputs. The framework aims to deliver the highest possible quality for users and is influenced by two primary cost factors: (i) placement costs, representing the energy and resource overhead of deploying services at edge nodes, and (ii) transmission costs, typically associated with latency requirements. Our system dynamically allocates GDM services to edge nodes while managing multi-user access to shared wireless channels. Our main contributions are outlined as follows:
\vspace{-4pt}
\begin{itemize}
    \item Unified optimization framework for joint channel allocation to users (for prompt/initial-condition transmission) and placement of GDM services on edge-computing-equipped BSs. Notably, GDMs are iterative and quality-progressive, produce large intermediate latents, and thus impose strict ordering, latency, and communication–compute coupling that require resource allocation.
    \item LEARN-GDM: a decision-making algorithm that partitions denoising blocks across heterogeneous nodes, incorporates latent transmission costs among them, and enables adaptive reduction of denoising steps to balance performance and resource consumption.
    \item Simulations demonstrating scalability and flexibility by evaluating (i) the impact of available access channels on user-request quality and (ii) quality maintenance under increasing numbers of simultaneous requests.
\end{itemize}

\vspace{-4pt}
In the remainder of this paper, Section \ref{PR_F} introduces the system model and formulates the optimization framework. Section \ref{S_SCHEME} details the proposed approach to service placement and multiple access control. The numerical results are presented in Section \ref{S_EVAL}, while final remarks and directions for future research are presented in Section \ref{S_CON}.

\section{Problem Definition}\label{PR_F}

\subsection{Glimpse into GDM}
A GDM service can be characterized as a combination of forward and reverse processes. The forward process maps the initial state, denoted by \( x_0 \), to a noise vector \( x_\mathcal{B} \) (that is sampled from a Gaussian prior)\footnote{The forward process is employed during the training phase and is not within the scope of this study \cite{10419041, du2024enhancing}.}. The reverse process denoises it to its original state ($x_\mathcal{B} \to x_{\mathcal{B}-1} \to \ldots \to x_0$), conditioned on a prompt or guiding signal \( c \) and utilizing learned transition kernels. This process, parameterized by \( \theta \), is 

\begin{figure}[t!]\centering
\vspace{0.1cm}
\includegraphics[width=3.0in]{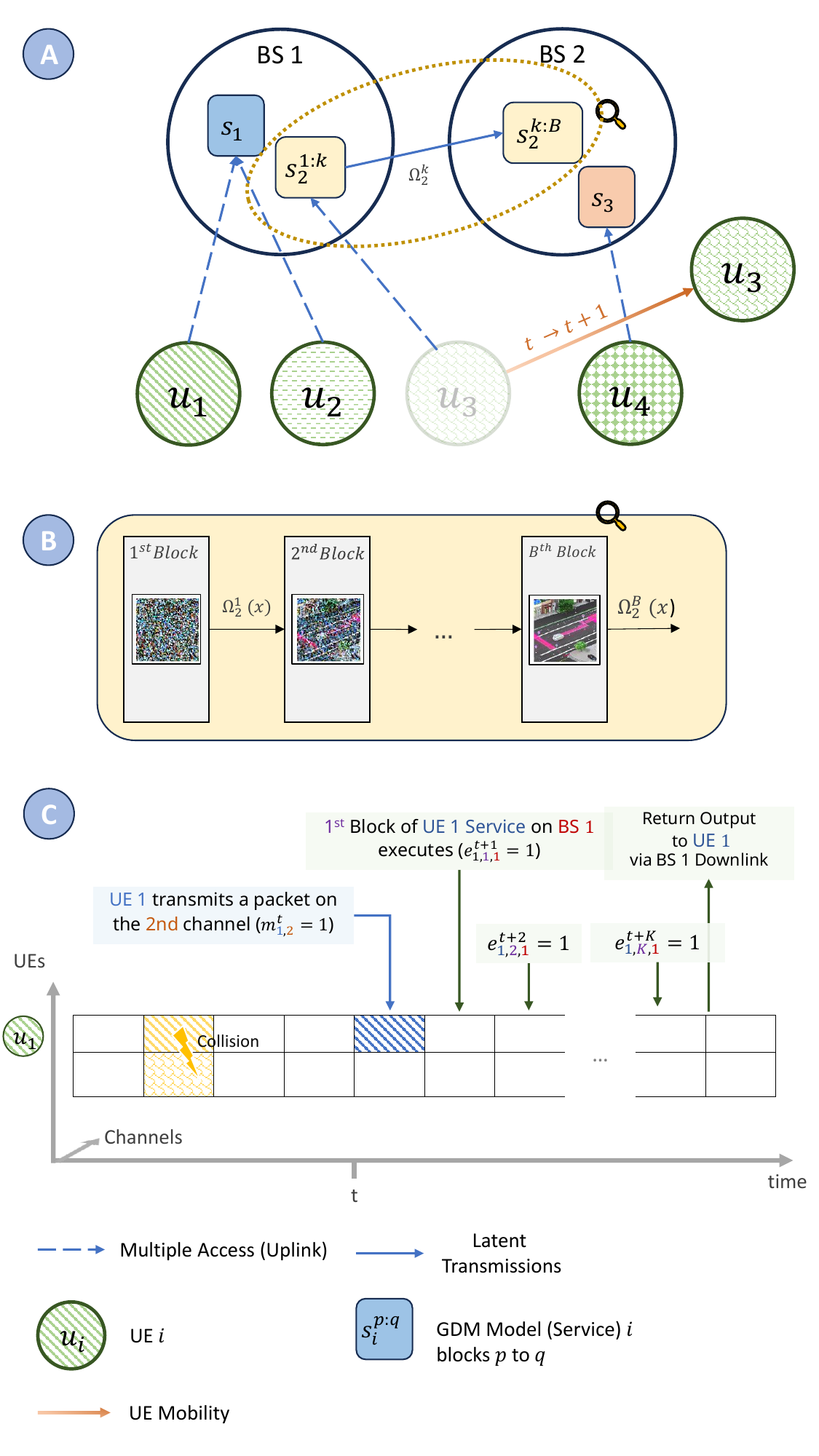}
\vspace{-0.5cm}
  \caption{Sample scenario with $\mathcal{N} = 2$, $\mathcal{U} = 4$ and $\mathcal{S} = 3$. UEs are fixed, except $u_{3}$, which changes its associated BS between two time frames. \textbf{A}) $u_{1}$ and $u_{2}$ are both registered to $s_{1}$, and $u_{3}$ and $u_{4}$ are registered to $s_{2}$ and $s_{3}$, respectively. \textbf{B}) GDM Model of $s_{2}$ with maximum $\mathcal{B}$ denoising step blocks. \textbf{C}) The first transmission attempt of $u_{1}$ resulted in a collision, while their next packets were transmitted successfully, resulting in the execution of the corresponding service. All blocks of $s_{1}$ are executed in BS $1$. On the other hand, the last $( \mathcal{B} - k)$ blocks of $s_{2}$ are executed in BS $2$ so that by generating the final synthetic data, $u_{3}$ is able to receive it from the associated BS.}
    \vspace{-0.55cm}
    \label{figure:systen_model}
\end{figure}

\footnotesize
\begin{align}\label{ddpm}
p_\theta(x_{t-1} \mid x_t, c) := \mathcal{N}(x_{t-1}; \mu_\theta(x_t, t, c), \Sigma_\theta(x_t, t, c)),
\end{align}
\normalsize
where \( \mu_\theta \) and \( \Sigma_\theta \) denote the learnable mean and variance of the reverse Gaussian transition kernel, respectively. Moreover, \( \mathcal{B} \) signifies the number of denoising step blocks, with \( \mathbbm{B} = \{ 1, \ldots, \mathcal{B} \} \) representing the set of their blocks, with ${\Omega}^{k}(x)$ output quality of the $k$-th block for input data $x$. Notably, a higher number of blocks provides more degrees of freedom for service placement, and yet enlarges the problem size.
In the example illustrated in Fig. \ref{figure:systen_model}-B, the output quality for various blocks is presented.

In this study, we assume that each block requires a single time frame for execution; however, this framework can be extended to scenarios in which the execution time of each block exceeds one time frame. Accordingly, it takes $\mathcal{B}$ time frames to fully complete a service, while executing only the first $K \le \mathcal{B}$ blocks of service is possible, aiming to trade off quality for resource efficiency. In addition, it is possible to execute different blocks in distinct BSs, based on available resources and UEs' mobility patterns.

\begingroup
\begin{table}[t!]
\caption{Notations of Symbols Used in the System Model.}
\vspace{-1em}
\setlength\tabcolsep{2.0pt}
\setlength\extrarowheight{-3pt}
\renewcommand{\arraystretch}{2.0}
\begin{center}
\small
\begin{tabular}{|c|l|}
\hline
\textbf{Symbol} & \multicolumn{1}{c|}{\textbf{Description}} \\ \hline
$u_{i}$ & UE with index $i \in \mathbbm{U} = \{1, \ldots, \mathcal{U}\}$ \\
$n$ & BS index $\in \mathbbm{N} = \{1, \ldots, \mathcal{N}\}$ \\
$s$ & GDM service index $\in \mathbbm{S} = \{1, \ldots, \mathcal{S}\}$ \\
$c$ & Communication channel index $\in \mathbbm{C} \! = \! \{1, \ldots, \mathcal{C}\}$ \\
$t$ & Time frame index $\in \mathbbm{T}$ \\
$\mathbbm{B}$ & Set of blocks in a service, $\{1, \ldots, \mathcal{B}\}$ \\
$\mathbbm{P} / \ \mathbbm{P}^{k}$ & Set of all / $k$-length execution paths \\
$\mathfrak{p}$ & A specific execution path $\in \mathbbm{P}$ \\
$\mathcal{J}^{k}_{\mathfrak{p}, n} \in \{ 0, 1 \}$ & Indicator if $n$ is used in path $\mathfrak{p}$ at step $k$ \\
$\Psi = [\psi^{t}_{i, n}]$ & Association graph: 1 if $u_i$ connected to $n$ at $t$ \\
$\Lambda = [\lambda_{i, s}]$ & UE to service assignment matrix \\
$r^{t}_{i, \mathfrak{p}} \in \{ 0, 1 \}$ & UE $i$ selects path $\mathfrak{p}$ at $t$ \\
$e^{t}_{i, k, n} \in \{ 0, 1 \}$ & \makecell[l]{block $k$ of UE $i$'s service, executed on BS $n$ at $t$} \\
$m^{t}_{i, c} \in \{ 0, 1 \}$ & UE $i$ uploads on channel $c$ at $t$ \\
$m^{t}_{i} \in \{ 0, 1 \}$ & UE $i$ uploads on any channel at $t$ \\
$W^{t}_{n} / \ \hat{W}_{n}$ & \makecell[l]{Actual / Max blocks executed on BS $n$ at $t$} \\
$\Omega_{s}(k)$ & \makecell[l]{Service $s$ output quality with $k$ blocks execution} \\
$Q^{t}_{i} / \ \overline{Q}_{i}$ & \makecell[l]{Received / Min required output quality for UE $i$} \\
$\hat{Y}_{n, n'}$ & Cost of transmission from node $n$ to $n'$ \\
$Y^{t}_{i}$ & Total transmission cost for UE $i$ at $t$ \\
$\epsilon_{n}$ & Execution cost of a single inference on BS $n$ \\
$\alpha, \beta$ & \makecell[l]{Execution and Transmission cost trade-off} \\ \hline
\end{tabular}
\label{tab_symbols}
\end{center}
\end{table}
\endgroup

\subsection{System Model}
The system consists of a set of heterogeneous edge-computing-equipped BSs, denoted with $\mathbbm{N} = \{ 1, \ldots, \mathcal{N} \}$, capable of providing GDM services to a collection of $\mathcal{U}$ mobile User Equipment (UE). This heterogeneity arises from various BS types, such as roadside units (RSUs), macro, and micro nodes, each with distinct computational capacities. UEs are uniquely labeled as $u_{i}$, where $i \in \mathbbm{U} = \{1, \ldots, \mathcal{U}\}$.
Fig. \ref{figure:systen_model} depicts a modest scenario with $\mathcal{N} = 2$ and $\mathcal{U} = 4$. To execute each service with \(k\) blocks, it is necessary to specify its \textit{execution path}, which is defined as the sequence of \(k\) nodes \( ( {n_{1}, \ldots, n_{k}} ) \), with each node representing the execution BS of its corresponding block. This sequence is denoted by \(\mathfrak{p} \in \mathbbm{P} = \bigcup_{k \in \mathbbm{B}}{\mathbbm{P}^{k}}\), where \(\mathbbm{P}^{k}\) represents the set of all \(k\)-permutations of \(\mathbbm{N}\) with repetitions permitted\footnote{Evidently, this set is composed of $\mathcal{N}^{k}$. For example, even in a small scenario with only four nodes and five blocks, ${4}^{5} = 1024$ paths exist. Therefore, a subset of existing paths should be considered in practice.}. In this regard, the inclusion of node $n$ in path $\mathfrak{p}$ at step $k$ is represented with $\mathcal{J}^{k}_{\mathfrak{p}, n}$. In the scenario of Fig. \ref{figure:systen_model} with $\mathcal{B} = 2$, set $\mathbbm{P}^{2} = \{ (1, 1), (1, 2), (2, 1), (2, 2) \}$, where the first path executes two blocks on BS $1$ and the second path executes on BS $1$ and $2$ respectively. For the second path (i.e., $(1, 2)$), $\mathcal{J}^{1}_{2, 1}$ and $\mathcal{J}^{2}_{2, 2}$ are equal to 1. Table \ref{tab_symbols} consolidates the aforementioned symbols and the others discussed hereafter.

The system is divided into a total of $\mathcal{A}$ predefined service areas, which vary in size due to geographical factors such as obstacles. UEs move dynamically between areas, but for simplicity, they are assumed to stay within a single area \textit{during} each time frame $t \in \mathbbm{T}$. These mobilities create a dynamic association graph $\Psi = [\psi^{t}_{i, n}]_{\mathcal{U} \times \mathcal{N} \times \mathcal{T}}$, equals 1 if $u_{i}$ is connected to BS $n$ at time frame $t$, otherwise 0. In the example of Fig. \ref{figure:systen_model}, the association graph $\Psi^{t} = \big[ [1, 0], [1, 0], [1, 0], [0, 1] \big]$. UEs in the same area are engaged in contention for access to a set of $\mathbbm{C} = \{1, \ldots, \mathcal{C}\}$ perfectly time-slotted communication channels designated for uploading their data to BSs. Consequently, their simultaneous transmissions over a single channel lead to a collision. In contrast, service responses are delivered to UEs via collision-free downlink channels. It is also assumed that BSs have broadband channels between them.

The system offers a collection of GDM services represented by the set $\mathbbm{S} = \{1, \ldots, \mathcal{S}\}$. We consider each service $s$ as a trained and ready-to-use GDM model. UEs have been assigned to services through a predefined matrix $\Lambda = [\lambda_{i, s}]_{\mathcal{U} \times \mathcal{S}}$. This matrix is constructed through a scalable and dynamic service discovery mechanism \cite{farhoudi2024discovery}. In the example of Fig. \ref{figure:systen_model}, matrix $\Lambda = \big[ [1, 0, 0], [1, 0, 0], [0, 1, 0], [0, 0, 1] \big]$.

\subsection{Problem Formulation}
To fulfill the services requested by UEs, the initial step is to determine the execution path for each UE.

The binary decision variable ${r}^{t}_{i, \mathfrak{p}}$ indicates selection of execution path $\mathfrak{p}$ for $u_{i}$ at the beginning of time frame $t$. Moreover, the support variable ${e}^{t}_{i, k, n}$ is responsible for indicating whether block $k$ of UE $i$'s service is executed on the BS $n$ at time frame $t$. The variables \( r \) and \( e \) are subject to constraint C1, which specifies that the selection of a path of length \( k \) necessitates the execution of all \( k \) blocks of the associated service at the corresponding nodes over the subsequent \( k \) time frames. Hereafter, we assume out-of-bounds indices are 0 for simplicity. Moreover, each UE can select only one path per time frame (C2), and each BS $n$ can execute maximum $\hat{W}_{n}$ blocks per time frame (C3)\footnote{While all blocks of a GDM service are implemented with the same neural network, they require separate inferences and thus separate processing power.}.


\vspace{-10 pt}
\footnotesize
\begin{align}\label{e_r_constraints}
& {r}^{t}_{i, \mathfrak{p}} \le \!\!\!\! \ \ \sum_{n \in \mathbbm{N}} \!\!\! \ \ {e}^{t+k-1}_{i, k, n}  \cdot  \mathcal{J}^{k}_{\mathfrak{p}, n} \quad  \forall i \in \mathbbm{U}, \mathfrak{p} \in \mathbbm{P}, k \in \mathbbm{B}, t \in \mathbbm{T} \tag{C1} \\
& \sum_{\mathfrak{p} \in \mathbbm{P}}^{}{{r}^{t}_{i, \mathfrak{p}} \le 1} \qquad \forall i \in \mathbbm{U}, \forall t \in \mathbbm{T}
\tag{C2} \\
& {W}^{t}_{n} \triangleq \makecell{\sum_{i \in \mathbbm{U}, k \in \mathbbm{B}} {e}^{t}_{i, k, n} \le \hat{W}_{n} \qquad \forall n \in \mathbbm{N}, t \in \mathbbm{T}} \tag{C3}
\end{align}
\normalsize


Another binary decision variable is ${m}_{i, c}^{t}$, which indicates the upload transmission of UE $i$ on channel $c$ at time frame $t$. Each UE can transmit only on one channel, and no more than one of the UEs connected to the same BS may transmit data simultaneously, as stated by channel constraints in C4 and C5, respectively.

\vspace{-10 pt}
\footnotesize
\begin{align}\label{m_constraints_1}
& {m}_{i}^{t} \triangleq \sum_{c \in \mathbbm{C}}^{}{{m}_{i, c}^{t}} \leq 1 \qquad \forall i \in \mathbbm{U}, t \in \mathbbm{T} \tag{C4} \\
& \sum_{i \in \mathbbm{U}}^{}{{m}_{i, c}^{t} \cdot \psi^{t}_{i, n}} \leq 1 \qquad \forall n \in \mathbbm{N}, c \in \mathbbm{C}, t \in \mathbbm{T} \tag{C5}
\end{align}
\normalsize

In order to establish an end-to-end transmission for each UE and its designated execution path, it is essential to correlate the variables \( e \) and \( m \). Constraint C6 facilitates this linkage by stipulating that the initial block of the corresponding service may only be executed if a request is present in the preceding time frame \( t-1 \). Recall that the initial block starts the denoising process from noise, with the UE prompt serving as a required conditioning input.

\vspace{-10 pt}
\footnotesize
\begin{align}\label{m_e}
\makecell{\sum_{\mathfrak{p} \in \mathbbm{P}}{r}^{t}_{i, \mathfrak{p}} \leq \sum_{c \in \mathbbm{C}}{{m}_{i, c}^{t-1} } \qquad \forall i \in \mathbbm{U}, t \in \mathbbm{T}} \tag{C6}
\end{align}
\normalsize
Moreover, the end-to-end assignment of each UE must be ensured in terms of output quality. We define in C7 the quality of synthetic data generated for UE $i$ at time frame $t$ as the number of executed blocks applied to function $\Omega_{s}(.)$, which generally increases with $k$. In practice, a minimum acceptable quality is set, represented by $\overline{Q}$ in C8. For example, in Fig. \ref{figure:SSIM_sample} if $\overline{Q} = 0.5$, delivering the output at step 10 offers no advantage. Notably, C8 must be satisfied if at least one path is selected for the UE.

\vspace{-10 pt}
\footnotesize
\begin{align}\label{qos}
{Q}^{t}_{i} & \triangleq \!\!\!\! \sum_{s \in \mathbbm{S}, \mathfrak{p} \in \mathbbm{P}} \!\!\! {{r}^{t}_{i, \mathfrak{p}} \cdot \lambda_{i, s} \cdot \Omega_{s}(|\mathfrak{p}|)} \tag{C7} \\
&\geq \overline{Q}_{i} \cdot \sum_{\mathfrak{p} \in \mathbbm{P}}^{}{r}^{t}_{i, \mathfrak{p}} \qquad \forall i \in \mathbbm{U}, t \in \mathbbm{T} \tag{C8}
\end{align}
\normalsize

The final step is to establish a metric to measure the costs of intermediate latent data transfer between executing nodes, along with the transfer of UE data from the Point of Attachment (PoA) to the first execution node and the generated data from the last execution node to the final PoA (C9). Note that in this formulation, the execution path head and tail can differ from PoA nodes, but their negative impact on the transmission cost is considered. Here, $\hat{Y}_{n, n'}$ is the cost of transmitting data from node $n$ to $n'$.

\vspace{-10 pt}
\footnotesize
\begin{align} \label{cost}
{Y}^{t}_{i} = & \sum\limits_{ \makecell{{ \mathfrak{p}, (n, n'), k} \\ \in \mathbbm{P}, \mathbbm{N} \times \mathbbm{N}, \mathbbm{B}}}^{}{ \hspace{-2em} {r}^{t}_{i, \mathfrak{p}} \cdot \mathcal{J}^{k}_{\mathfrak{p}, n} \cdot \mathcal{J}^{k+1}_{\mathfrak{p}, n'} \cdot \hat{Y}_{n, n'} } \quad \notag \\
& + \sum\limits_{ \makecell{{ (n, n'), \mathfrak{p}} \\ \in \mathbbm{N} \times \mathbbm{N}, \mathbbm{P}}} {\hspace{-2em} {r}^{t}_{i, \mathfrak{p}} \Big(  \psi^{t-1}_{i, n} \cdot \mathcal{J}^{1}_{\mathfrak{p}, n'} + \psi^{t+|\mathfrak{p}|}_{i, n} \cdot \mathcal{J}^{|\mathfrak{p}|}_{\mathfrak{p}, n'} \Big) \cdot \hat{Y}_{n, n'} } \notag
\\ 
&\forall i \in \mathbbm{U}, t \in \mathbbm{T} \tag{C9}
\end{align}
\normalsize

Now, we can set the objective of our problem to maximize the total quality of UEs over all time frames, subtracted by the scaled total cost of service execution and data transfer. Here, $\epsilon_{n}$ is the cost of a single inference on node $n$. Given that this objective exhibits non-linearity and certain parameters, such as $\Omega_{s}(.)$ (i.e., the quality per block function), are not practically known, traditional gradient-based optimization techniques may struggle to converge to an optimal solution.

\vspace{-10 pt}
\footnotesize
\begin{align} \label{problem}
&\mathrm{ max } \sum_{i \in \mathbbm{U}, t \in \mathbbm{T}}^{}{{Q}^{t}_{i}} - \alpha \cdot \sum_{n \in \mathbbm{N}, t \in \mathbbm{T}}{\epsilon_{n} \cdot {W}^{t}_{n}} - \beta \cdot \sum_{i \in \mathbbm{U}, t \in \mathbbm{T}}^{}{{Y}^{t}_{i}} \notag \\ 
& \text{s.t. (C1-C9) } 
\end{align}
\normalsize

\section{Proposed Scheme}\label{S_SCHEME}

To address the previously mentioned challenges associated with problem complexities and incomplete knowledge, we propose muLtiplE Access and seRvice placemeNt for GDMs (LEARN-GDM), an intelligent resource allocation algorithm that allocates channels and places services per time frame.
LEARN-GDM operates effectively despite uncertain system information since it incorporates user mobility prediction and proactively deploys services closer to where users are likely to move. 

LEARN-GDM consists of a greedy Multiple Access Algorithm (MAC) and Double and Dueling Deep Q-Learning (D3QL)-based service placement \cite{mazandarani2023self, mazandarani2024semantic, deeplearningbasedservice}. Deep Reinforcement Learning, including D3QL as a value-based method, has been extensively utilized to address complex network optimization challenges, even with limited knowledge of wireless networks, while continuously learning and enhancing understanding of the environment \cite{alwarafy2022frontiers}.

The MAC component in LEARN-GDM greedily assigns channels to UEs connected to the same BS, prioritizing those whose ongoing inference processes are closest to the quality threshold. For example, between two UEs with ongoing qualities $0.3$ and $0.4$, the UE with $0.4$ has higher priority when the threshold is $0.5$; if the threshold is $0.25$, both would have equal priority.

D3QL enhances Deep Q-Learning (DQL) by decoupling action selection and evaluation \cite{hasselt_deep_2016} using the target value defined in \eqref{eq_DDQL_target}, followed by integrating Wang \textit{et al.}'s dueling approach \cite{wang2016dueling}. This target incorporates reward $\rho$, observation $O$, real action $a \in \boldsymbol{\mathcal{A}}$, and predicted action $a'$ to update DQL weights ($\mathcal{W}$) for each observation-action of time $t$ ($O^t, a^t$). Here, $a'\!\!\!=\!\! \text{argmax}_{a \in \boldsymbol{\mathcal{A}}} Q({O}^{t+1}, a; \mathcal{W}^t)$ with $\mathcal{W}$ representing evaluation weights updated at each step, and $\mathcal{W}^-$ as target network weights, synchronized every $\hat{t}\!\gg\!0$ step. Moreover, separate estimators calculate state values ($\mathcal{V}$) and action advantages ($\mathcal{AD}$), combining them to compute Q-values \eqref{eq_dueling} and weights \eqref{eq_update_w}. This approach improves training stability, accelerates convergence, and mitigates overestimation issues.

\footnotesize
\vspace{-5pt}
\begin{equation}
\begin{aligned}
Y^t  = {\rho}^{t} + \gamma \cdot Q({O}^{t+1}, a'; {\mathcal{W}}^{t-}) \hspace{15em}
\end{aligned}
\label{eq_DDQL_target}
\end{equation}
\vspace{-3mm}
\normalsize

\footnotesize
\vspace{-23pt}
\begin{align}
\label{eq_dueling}
Q(O^t, a^t; \mathcal{W}^t) =  \mathcal{V}(O^t; \mathcal{W}^t) \nonumber + \Bigg(  & \mathcal{AD}(O^t, a^t; \mathcal{W}^t) \nonumber \hspace{4em} \\
&- \frac{1}{\left| \boldsymbol{\mathcal{A}} \right|} \sum_{a^{\prime} \in \boldsymbol{\mathcal{A}}}^{}\mathcal{AD}(O^t, a^{\prime}; \mathcal{W}^t) \Bigg)
\end{align}
\normalsize

\footnotesize
\vspace{-13pt}
\begin{equation}
\begin{aligned}
     \mathcal{W}^{t+1} \gets \mathcal{W}^{t} + \sigma \cdot [Y^t - Q(O^t, a^t; \mathcal{W}^t)] \nabla_{\mathcal{W}^t} Q(O^t, a^t; \mathcal{W}^t) \hspace{2em}
\end{aligned}
\label{eq_update_w}
\end{equation}
\vspace{-3mm}
\normalsize

To define action space ($\boldsymbol{\mathcal{A}}$), it determines which BS (if any) should initialize or continue a process for the next time frame. A non-zero action for UE $i$ signifies its continuity if it has already started but not yet finished. More concretely, if block $k < \mathcal{B}$ is executed in the current time frame, a non-zero value indicates that block $k+1$ will be executed in the next time frame; otherwise, the first block will be executed. Conversely, a zero value prevents starting a new inference process or stops an existing service. When a service stops, the generated image is delivered to the corresponding UE.

\vspace{-10 pt}
\footnotesize
\begin{align} \label{action}
\boldsymbol{\mathcal{A}} = \big\{ a_{i}: \emptyset \cup \mathbbm{N} \ | \ \forall i \in \mathbbm{U} \big\}
\end{align}
\normalsize
Observation space must provide the resource allocator with sufficient information regarding the current and historical state of the allocated resources, as well as the ongoing state of each chain and multiple access information.

\vspace{-10 pt}
\footnotesize
\begin{align} \label{state}
{o}^{t} &= \makecell[l]{\\ \big\{ {W}_{n}/\hat{W}_{n}, \epsilon_{n} | n \in \mathbbm{N} \big\} \cup 
\big\{ {Q}^{t}_{i} - \overline{Q}_{i} | i \in \mathbbm{U} \big\} \cup
\big\{ {m}_{i}^{t - 1} | i \in \mathbbm{U} \big\} \\ 
\ \cup ~
\big\{\psi^{t}_{i, n} | n \in \mathbbm{N}, i \in \mathbbm{U} \big\} } \notag
\\
{O}^{t} &= \Big\{ {o}^{h} | h \in \{ t-{\mathcal{H}}, \ldots, t \} \Big\}
\end{align}
\normalsize
To circumvent the problem of delayed rewards \cite{li2024hype}, we reward users based on their image quality increase resulting from successful allocations and constrained to exceed the quality threshold, subtracted by the scaled cost of allocations and transmissions.

\vspace{-10 pt}
\footnotesize
\begin{align}\label{reward}
{\rho}^{t} = \makecell[l]{\\ \sum_{i \in \mathbbm{U}}^{}{ \big( ( {Q}^{t}_{i} - {Q}^{t-1}_{i} ) \cdot \mathbbm{1}({Q}^{t}_{i} \ge \overline{Q}_{i} ) \big)  } \\ - ~ \alpha \cdot \big( \sum_{n \in \mathbbm{N}}{\epsilon_{n} \cdot {W}^{t}_{n}} \big) - \beta \cdot \big( \sum_{i \in \mathbbm{U}}^{}{{Y}^{t}_{i}} \big)}
\end{align}
\normalsize

Our approach, outlined in Algorithm \ref{alg_LEARN_GDM}, consists of several key components. Steps 4–8 implement the greedy MAC algorithm, while steps 10–14 apply an epsilon-greedy action selection strategy for service placement. For each UE, the placement procedure (steps 16–21) is executed as follows: the system first checks whether the maximum chain length has been reached. If it has, the inference process concludes, and the result is delivered to the UE. If not, the placement decision depends on the selected action--specifically, whether the action assigns a service placement to the UE--and the current node capacity status. If inference has already been initiated, the placement advances to the next block; if not, it begins at the first block. In cases where the action is null or the node's capacity is exhausted, any available latent data is sent to the UE. Finally, steps 23–28 address the D3QL agent training phase.
\vspace{-5pt}

\begin{algorithm}[t!]\label{alg_LEARN_GDM}
\caption{\small{muLtiplE Access and seRvice placemeNt for GDMs (LEARN-GDM)}}
\KwInput{$\mathcal{T}$, Set of Episodes ($\mathbb{E}$), }
${\mathcal{W}} \leftarrow \mathbf{0}$, ${\mathcal{W}}^{-} \leftarrow \mathbf{0}$, $\epsilon \gets 1$, $memory \gets \{\}$ \\
\ForEach{$ep$ in $\mathbb{E}$}
{
    \ForEach{$t$ in $\mathbb{T}$}
    {   
        \vspace{5 pt} \textcolor{gray}{$\star$ Multiple Access $\star$}  \\
        UE Priorities $\gets \max{ \Big\{ 1 / ( \overline{Q} - {Q}^{t} ), 10^{-8} \Big\} }$ \\
        ${\mathbbm{U}}_{sorted} \gets \textit{SORT}_{u} \big\{ {\mathbbm{U}}, \textit{key}= \small\text{UE Priorities} \big\}$ \\
        \ForEach{$c$, $i$ in ${\mathbbm{C}}, {\mathbbm{U}}_{sorted}[1:\mathcal{C}]$}
        {
            ${m}_{i, c}^{t} \gets 1 $
        }
        \vspace{5 pt} \textcolor{gray}{$\star$ Service Placement $\star$}  \\
        $\zeta \gets$ sample uniformly from $\textit{Uniform}(0, 1)$ \\
        \If{$\zeta > \epsilon$}
        {
            $ a^{t} \gets \text{argmax}_{a' \in \boldsymbol{\mathcal{A}}} Q({O}^{t+1}, a'; \mathcal{W}^t)$ \\
        }
        \Else
        {
            select a random $a^{t}$ from $\boldsymbol{\mathcal{A}}$
        }
        \vspace{5 pt} \ForEach{$i$ in ${\mathbbm{U}}_{sorted}$}
        {
            \vspace{2 pt} \If{Maximum blocks ($\mathcal{B}$) has reached}
            {
                Deliver the result to the UE
            }
            \vspace{2 pt} \ElseIf{($a_{i}^{t} = n \in \mathbbm{N}) \wedge (W_n < \hat{W}_n)$}
            {
                Deploy the first/next block on $n$ \\
            }
            \Else
            {
                Deliver the result (if available) to UE
            }
        }
        \vspace{2 pt} Observe ${\rho}^{t}$ and construct ${O}^{t+1}$ acc. to \eqref{state} \\
        \vspace{5 pt} \textcolor{gray}{$\star$ Training $\star$} \\
        $memory \gets memory \cup \{(\rho^{t}, {O}^{t}, a^{t}, {O}^{t+1}) \}$ \\
        Choose a batch of samples from $memory$\\
        Train the agent according to \eqref{eq_update_w} \\
        \If{$\epsilon > \widetilde{\epsilon}$}
        {
            $\epsilon \gets \epsilon \cdot \epsilon'$
        }
    }
}
\end{algorithm}
\vspace{-10pt}

\section{Performance Evaluation}\label{S_EVAL}

\begin{figure}[t!]\centering
\vspace{0.2cm}
\includegraphics[width=3.2in]{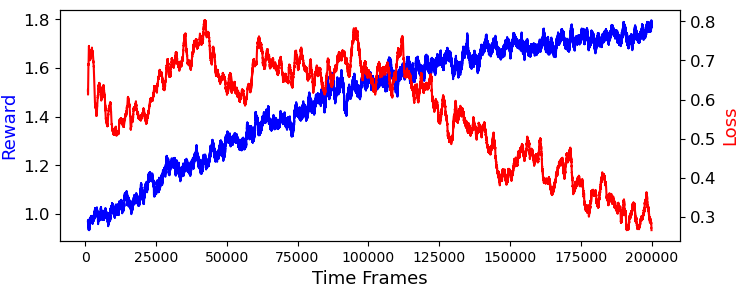}
\vspace{-0.4cm}
\caption{Reward and Mean Squared Error (MSE) Loss evolution of the service placement learning algorithm over 5,000 training episodes (each with 40 time frames), showing increased rewards and decreased MSE loss over time.}
\vspace{-0.5cm}
\label{figure:reward}
\end{figure}

In this section, we conduct a numerical analysis of the D3QL-based solution using the parameters outlined in Table \ref{t_prm}. As a first step, we analyze convergence, followed by a comparison of performance. To illustrate the learning process involved in service placement, Fig. \ref{figure:reward} displays the service placement reward (blue line) within learning episodes. The learning algorithms are trained over 200,000 time frames. The reward increases gradually, which illustrates the efficiency of DRL in placement. Additionally, rewards stabilize after the 175,000 time frame, indicating convergence.

For comparison, we explore two practical scenarios: 1) the impact of user numbers on scalability; 2) how the number of channels, indicative of a communications bottleneck in real-world applications, affects performance. Notably, in all scenarios, UEs are randomly distributed across the grid and move according to the Random Waypoint Model, with an average speed of 10 m/s and a pause time of 3 seconds. During comparison, we employ four benchmarks. First, the Monolithic Placement (MP) method, indicated by a purple line, operates with a single node per inference, placing a flexible number of blocks on that node. This method is a relaxed version of the approaches proposed by \cite{liu2025qos}, representing an upper bound for their results in the present analysis. Second, a Fixed Chain Placement (FP), which is also based on the D3QL algorithm, lacks the flexibility of variable chain lengths that are helpful for tradeoffs. Third, a Greedy algorithm (GR) illustrated by a red line serves as another benchmark by assigning each block to the PoA. Finally, the Optimization algorithm (OPT) solves the problem defined in \eqref{problem} with full knowledge, utilizing the Gurobi optimization solver \cite{gurobi}, and establishes a universal upper bound applicable to all approaches.

\begin{table}[t!]
\caption{System Model Parameters.}
\begin{center}
\begin{tabular}{|c|c|}
\hline
\textbf{Parameter} & \textbf{Value} \\
\hline
Network area & 4x4 grid \\
Node Processing Capacity ($\hat{W}$) & $\sim \mathcal{U}(1, 3)$ \\
Node Placement cost ($\epsilon$) & $\sim \mathcal{U}(1, 4)$ per inference \\
Quality Threshold (${\overline{Q}}$) & $\sim \mathcal{U}(0.1, 0.5)$ \\
Number of Services (${\mathcal{S}}$) & 3 \\
Max. blocks per service (${\mathcal{B}}$) & 4 \\
Default number of UEs & $15$ \\
Default number of channels & $2$ \\
Scaling Factors ($\alpha$, $\beta$) & $0.1$, $0.1$ \\
LSTM history size ($\mathcal{H}$) & $3$ experiences \\
Capacity of experience memory & $5000$ experiences \\
Batch size & $32$ \\
Discount factor ($\gamma$) & 0.9 \\
Learning rate & $0.0008$ \\
Exploration parameters $\widetilde{\epsilon}$, $\epsilon'$ & 0.00001, 0.99995 \\
Approximator model & \begin{tabular}{@{}c@{}c@{}} LSTM with 128 units \\ + fully-connected layers \\ with $128$, $64$ and $32$ units \end{tabular}  \\
\begin{tabular}{@{}c@{}} Target network update frequency \\ \end{tabular} & Every $150$ steps \\
\hline
\end{tabular}
\label{t_prm}
\end{center}
\vspace{-10pt}
\end{table}

\subsubsection{Number of Users}

The number of users reflects the scalability of the system concerning growing demand. In this scenario, we put this under test by setting different values of \(\mathcal{U}\). Fig. \ref{figure:aggregated}-(A) clearly shows superiority of our approach over MP, FP, and GR methods. OPT possesses knowledge of the UE movements; therefore, they keep their performance despite heavy loads. Among MP and FP, the comparison depends on the problem scale. In a small number of UEs, MP outperforms, meaning that in such situations, having the option of variable chain length matters more. On the other hand, in a large number of UEs, distributing diffusion blocks on various nodes yields more benefit than setting a variable chain length.

\subsubsection{Number of Channels}
\begin{figure}[t!]\centering
\vspace{0.2cm}
\includegraphics[width=3.4in]{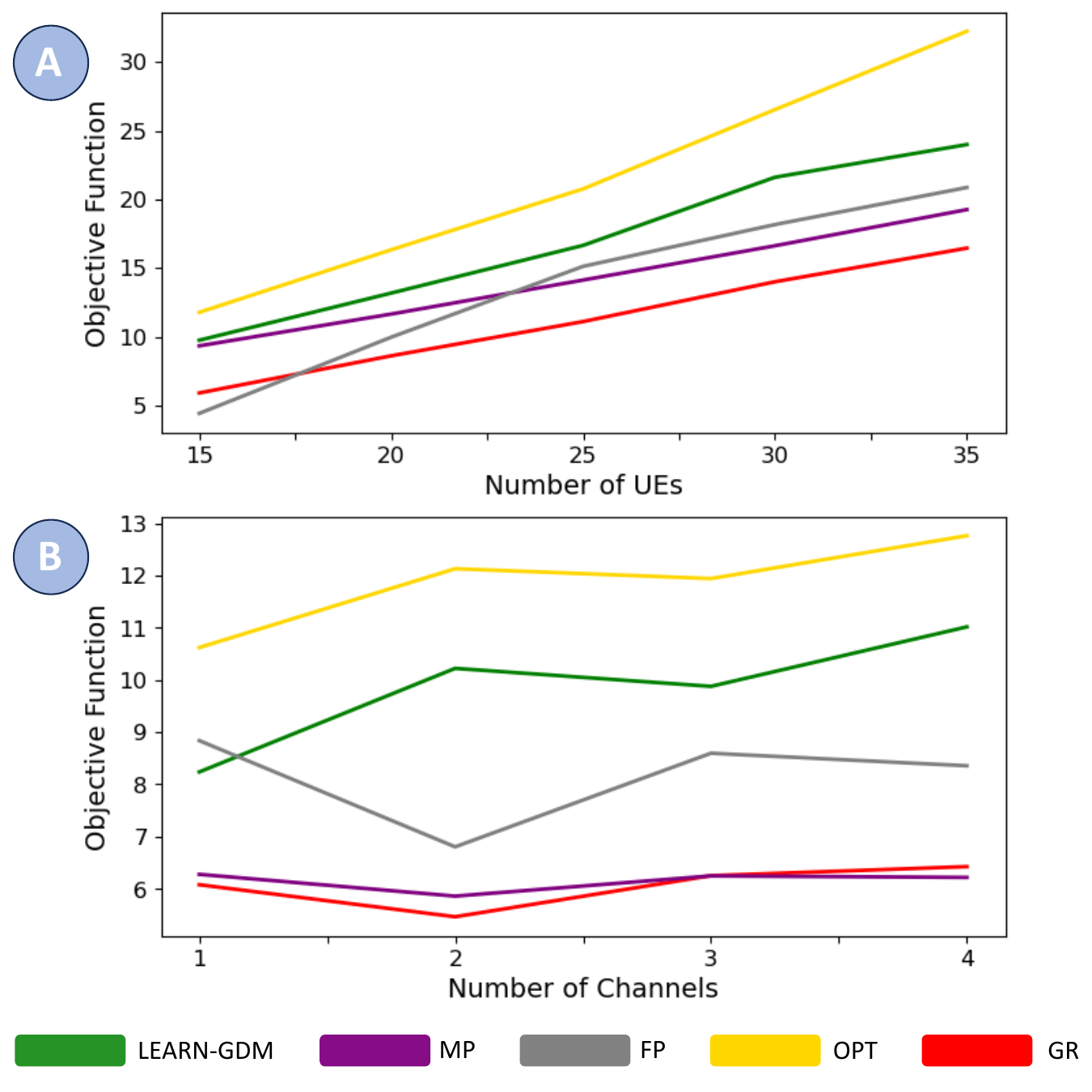}
\vspace{-0.1cm}
\caption{Performance comparison of LEARN-GDM against baseline methods: Monolithic Placement (MP), Fixed Chain Placement (FP), Optimization algorithm (OPT), and Greedy algorithm (GR) under varying system settings where (A) the number of UEs increases. (B) The number of channels increases.}
\label{figure:aggregated}
\end{figure}

With the proliferation of GenAI services and in particular, GDMs on the edge, it is expected that the communications factors of the system play an important role in service provisioning. By varying the number of available channels (denoted as \(\mathcal{C}\)), we can assess how the communications bounds affect system performance. As illustrated in Fig. \ref{figure:aggregated}-(B), a reduced number of channels leads to slightly increased collisions, which hinders the users' capacity to transmit their data containing prompts and initial conditions to the executing nodes. However, our approach demonstrated a considerably diminished negative impact in comparison to other methods. This resilience is attributed to its flexibility of variable chain lengths and executing nodes, the lack of which makes the MP, FP, and GR methods struggle.

\section{Conclusion}\label{S_CON}
This work introduced a unified problem formulation for channel allocation to GenAI users for transmitting their prompts or conditioning inputs, alongside the placement of GDM services on edge-computing-enabled BSs. The problem considered the distribution of denoising blocks across multiple nodes, accounting for latent transmission costs, and potentially reducing the number of denoising steps to balance performance with resource consumption. Moreover, we proposed and evaluated LEARN-GDM, a decision-making algorithm built upon D3QL, which is an enhanced version of Deep Q-Learning. To this end, the state and action spaces, as well as the reward mechanism, were thoroughly customized for the problem at hand. Our analysis demonstrates that the proposed algorithm enhanced overall QoS compared to conventional approaches.

While this study focuses on GDMs, the proposed framework is modifiable to other gradual inference processes, such as DNN partitioning use cases \cite{DNNSplit} or Large Language Models (LLMs) deployments over space-air-ground integrated networks comprising numerous heterogeneous nodes \cite{shokrnezhad2025autonomous}. Additionally, our research aligns with the ongoing semantic revolution in communication systems, which emphasizes the role of data semantics in achieving task-specific goals \cite{our_mag_paper}. Future work includes extending this approach to semantic networking that enables co-inference—sharing computation blocks across GenAI services for different users. This can be facilitated by incorporating tasks with similar intents into a unified knowledge graph \cite{xie2024gai}. Last but not least, there is potential to explore GDM resource allocation in a quantum internet in the future, yielding greater sustainability and applicability \cite{prados2023deterministic}.

\section*{Acknowledgment}
\vspace{-5pt}
This work is partially conducted at ICTFICIAL Oy. It is partially supported by the European Union’s Horizon Europe programme for Research and Innovation through the 6G-SANDBOX project (Grant No. 101096328) and the 6G-Path project (Grant No. 101139172). The studies of  M. Farhoudi are partially supported by the Business Finland 6Bridge 6Core project (Grant Number 8410/31/2022). The paper reflects only the authors’ views. The European Commission and the Spanish Ministry are not responsible for any use that may be made of the information it contains.  

\bibliographystyle{IEEEtran}
\bibliography{IEEEabrv, conf_short, main}

\begin{thebibliography}{10}
\providecommand{\url}[1]{#1}
\csname url@samestyle\endcsname
\providecommand{\newblock}{\relax}
\providecommand{\bibinfo}[2]{#2}
\providecommand{\BIBentrySTDinterwordspacing}{\spaceskip=0pt\relax}
\providecommand{\BIBentryALTinterwordstretchfactor}{4}
\providecommand{\BIBentryALTinterwordspacing}{\spaceskip=\fontdimen2\font plus
\BIBentryALTinterwordstretchfactor\fontdimen3\font minus \fontdimen4\font\relax}
\providecommand{\BIBforeignlanguage}[2]{{%
\expandafter\ifx\csname l@#1\endcsname\relax
\typeout{** WARNING: IEEEtran.bst: No hyphenation pattern has been}%
\typeout{** loaded for the language `#1'. Using the pattern for}%
\typeout{** the default language instead.}%
\else
\language=\csname l@#1\endcsname
\fi
#2}}
\providecommand{\BIBdecl}{\relax}
\BIBdecl

\bibitem{10419041}
H.~Cao \emph{et~al.}, ``A survey on generative diffusion models,'' \emph{IEEE Trans. Knowl. and Data Eng.}, vol.~36, no.~7, pp. 2814--2830, 2024.

\bibitem{letafati2024conditional}
M.~Letafati \emph{et~al.}, ``Conditional denoising diffusion probabilistic models for data reconstruction enhancement in wireless communications,'' \emph{{IEEE} Trans. on Mach. Learn. Commun. Netw.}, vol.~3, pp. 133--146, 2025.

\bibitem{Salar_2025_CVPR}
A.~Salar, Q.~Liu, Y.~Tian, and G.~Zhao, ``Enhancing facial privacy protection via weakening diffusion purification,'' in \emph{Proc. IEEE/CVF Conf. Computer. Vis. and Pattern Recognit.}, June 2025, pp. 8235--8244.

\bibitem{du2024enhancing}
H.~Du \emph{et~al.}, ``Enhancing deep reinforcement learning: A tutorial on generative diffusion models in network optimization,'' \emph{{IEEE} Commun. Surveys Tuts.}, vol.~26, no.~4, pp. 2611--2646, 2024.

\bibitem{luong2025diffusion}
N.~C. Luong \emph{et~al.}, ``Diffusion models for future networks and communications: A comprehensive survey,'' \emph{arXiv preprint arXiv:2508.01586}, 2025.

\bibitem{ye2024diffusion}
J.~Ye, S.~Gong, L.~Chen, L.~Zheng, J.~Gao, H.~Shi, C.~Wu, X.~Jiang, Z.~Li, W.~Bi \emph{et~al.}, ``Diffusion of thought: Chain-of-thought reasoning in diffusion language models,'' \emph{Adv. Neural Information Process. Syst.}, vol.~37, pp. 105\,345--105\,374, 2024.

\bibitem{xie2024gai}
G.~Xie \emph{et~al.}, ``{GAI-IoV}: Bridging generative {AI} and vehicular networks for ubiquitous edge intelligence,'' \emph{{IEEE} Trans. Wireless Commun.}, vol.~23, no.~10, pp. 12\,799--12\,814, 2024.

\bibitem{tang2024hart}
H.~Tang \emph{et~al.}, ``{HART}: Efficient visual generation with hybrid autoregressive transformer,'' 2024.

\bibitem{feng2024exploring}
W.~Feng \emph{et~al.}, ``Exploring collaborative diffusion model inferring for {AIGC}-enabled edge services,'' \emph{{IEEE} Trans. on Cogn. Commun. Netw.}, pp. 1--1, 2024.

\bibitem{yang2024efficient}
W.~Yang, Z.~Xiong, S.~Guo, S.~Mao, D.~I. Kim, and M.~Debbah, ``Efficient multi-user offloading of personalized diffusion models: A {DRL}-convex hybrid solution,'' \emph{IEEE Trans. Mobile Comput.}, 2025.

\bibitem{zeng2025generative}
W.~Zeng \emph{et~al.}, ``Generative {AI}-aided multimodal parallel offloading for {AIGC} metaverse service in {IoT} networks,'' \emph{{IEEE} Internet Things J.}, pp. 1--1, 2025.

\bibitem{liu2025qos}
Y.~Liu \emph{et~al.}, ``{QoS}-aware multi-{AIGC} service orchestration at edges: An attention-diffusion-aided {DRL} method,'' \emph{{IEEE} Trans. on Cogn. Commun. Netw.}, pp. 1--1, 2025.

\bibitem{li2024flexgen}
P.~Li \emph{et~al.}, ``Flexgen: Efficient on-demand generative {AI} service with flexible diffusion model in mobile edge networks,'' \emph{{IEEE} Trans. on Cogn. Commun. Netw.}, pp. 1--1, 2024.

\bibitem{mazandarani2023self}
H.~Mazandarani, M.~Shokrnezhad, T.~Taleb, and R.~Li, ``Self-sustaining multiple access with continual deep reinforcement learning for dynamic metaverse applications,'' in \emph{Proc. IEEE Int. Conf. Metaverse Comput., Netw. and Appl. (MetaCom)}.\hskip 1em plus 0.5em minus 0.4em\relax IEEE, 2023, pp. 65--70.

\bibitem{shokrnezhad2025autonomous}
M.~Shokrnezhad and T.~Taleb, ``An autonomous network orchestration framework integrating large language models with continual reinforcement learning,'' \emph{{IEEE} Commun. Mag.}, vol.~63, no.~8, pp. 78--84, 2025.

\bibitem{shokrnezhad2024fairness}
M.~Shokrnezhad, H.~Mazandarani, and T.~Taleb, ``Fairness-utilization trade-off in wireless networks with explainable kolmogorov-arnold networks,'' in \emph{Proc. IEEE Virtual Conf. Commun. (VCC)}.\hskip 1em plus 0.5em minus 0.4em\relax IEEE, 2024, pp. 1--6.

\bibitem{mazandarani2025novel}
H.~Mazandarani, M.~Shokrnezhad, and T.~Taleb, ``A novel multiple access scheme for heterogeneous wireless communications using symmetry-aware continual deep reinforcement learning,'' \emph{IEEE Trans. Mach. Learn. Commun. and Netw.}, 2025.

\bibitem{shokrnezhad2024towards}
M.~Shokrnezhad \emph{et~al.}, ``Towards a dynamic future with adaptable computing and network convergence {(ACNC)},'' \emph{{IEEE} Netw.}, 2024.

\bibitem{farhoudi2024discovery}
M.~Farhoudi \emph{et~al.}, ``Discovery of {6G} services and resources in edge-cloud-continuum,'' \emph{{IEEE} Netw.}, vol.~39, no.~3, pp. 223--232, 2025.

\bibitem{mazandarani2024semantic}
H.~Mazandarani \emph{et~al.}, ``A semantic-aware multiple access scheme for distributed, dynamic {6G}-based applications,'' in \emph{Proc. IEEE Wireless Commun. and Networking Conf.}, 2024, pp. 1--6.

\bibitem{globecom2023}
M.~Farhoudi \emph{et~al.}, ``{QoS}-aware service prediction and orchestration in cloud-network integrated beyond {5G},'' in \emph{Proc. IEEE Global Telecommun. Conf.}, Dec. 2023, pp. 369--374.

\bibitem{deeplearningbasedservice}
------, ``Deep learning based service composition in integrated aerial-terrestrial networks,'' in \emph{IEEE Int. Conf. on Net. Soft.}, 2025, pp. 204--208.

\bibitem{alwarafy2022frontiers}
A.~Alwarafy, M.~Abdallah, B.~S. Ciftler, A.~Al-Fuqaha, and M.~Hamdi, ``The frontiers of deep reinforcement learning for resource management in future wireless hetnets: Techniques, challenges, and research directions,'' \emph{IEEE Open Journal of the Communications Society}, vol.~3, pp. 322--365, 2022.

\bibitem{hasselt_deep_2016}
H.~v. Hasselt, A.~Guez, and D.~Silver, ``\BIBforeignlanguage{en}{Deep reinforcement learning with double {Q}-learning},'' \emph{\BIBforeignlanguage{en}{Proc. of the AAAI Conference on Artificial Intelligence}}, vol.~30, no.~1, Mar. 2016.

\bibitem{wang2016dueling}
Z.~Wang \emph{et~al.}, ``Dueling network architectures for deep reinforcement learning,'' in \emph{Proc. Int. Conf. Mach. Learn.}, vol.~48, Jun. 2016, pp. 1995--2003.

\bibitem{li2024hype}
H.~Li \emph{et~al.}, ``From hype to reality: The road ahead of deploying {DRL} in {6G} networks,'' 2024.

\bibitem{gurobi}
\BIBentryALTinterwordspacing
{Gurobi Optimization, LLC}, ``{{Gurobi} {Optimizer} {Reference} {Manual}},'' 2023. [Online]. Available: \url{https://www.gurobi.com}
\BIBentrySTDinterwordspacing

\bibitem{DNNSplit}
P.~Kayal \emph{et~al.}, ``{DNNSplit}: Latency and cost-efficient split point identification for multi-tier {DNN} partitioning,'' \emph{{IEEE} Access}, vol.~12, pp. 80\,047--80\,061, 2024.

\bibitem{our_mag_paper}
M.~Shokrnezhad \emph{et~al.}, ``Semantic revolution from communications to orchestration for {6G}: Challenges, enablers, and research directions,'' \emph{{IEEE} Netw.}, 2024.

\bibitem{prados2023deterministic}
Prados-Garzon \emph{et~al.}, ``Deterministic {6GB}-assisted quantum networks with slicing support: A new {6GB} use case,'' \emph{{IEEE} Netw.}, vol.~38, no.~1, pp. 87--95, 2023.

\end{thebibliography}

\end{document}